\documentclass{aastex631}

\usepackage[normalem]{ulem}

\definecolor{ForestGreen}{HTML}{228B22}
\definecolor{deepgreen}{RGB}{0, 100, 0}

\begin{document}

\title{On the evidence of a dark matter density spike around the primary black hole in OJ~287}

\author[0000-0003-4067-5283]{Debabrata Deb}
\affiliation{The Institute of Mathematical Sciences, C.I.T. Campus, Taramani, Chennai 600113, India}

\author[0000-0003-4274-4369]{Achamveedu Gopakumar}
\affiliation{Department of Astronomy and Astrophysics, Tata Institute of Fundamental Research, Mumbai 400005, India}

\author{Mauri J. Valtonen}
\affiliation{FINCA, University of Turku, FI-20014 Turku, Finland}
\affiliation{Tuorla Observatory, Department of Physics and Astronomy, University of Turku, FI-20014 Turku, Finland}


\begin{abstract}
The central engine of blazar OJ~287 is arguably the most notable supermassive black hole (SMBH) binary candidate that emits nano-Hertz (nHz) gravitational waves. This inference is mainly due to our ability to predict and successfully monitor certain quasi-periodic doubly peaked high brightness flares with a period of $\sim$12 years from this blazer. The use of post-Newtonian accurate SMBH binary orbital description that includes the effects of higher order GW emission turned out to be a crucial ingredient for accurately predicting the epochs of such Bremsstrahlung flares in our SMBH binary central engine description for OJ~287. It was very recently argued that one should include the effects of dynamical friction, induced by certain dark matter density spikes around the primary SMBH, to explain the {\it observed} decay of SMBH binary orbit in OJ~287. Invoking binary pulsar timing-based arguments, measurements, and  OJ~287's orbital description, we show that observationally relevant SMBH binary orbital dynamics in OJ~287 are insensitive to dark matter-induced dynamical friction effects. This implies that we could only provide an upper bound on the spike index parameter rather than obtaining an observationally derived value, as argued by \cite{Chan2024}.
\end{abstract}


\keywords{BL Lacertae objects (158) --- Black hole physics (159) --- Dark matter density (354) ---Dark matter distribution (356)}


\section{Introduction} \label{sec:intro}

Individual SMBH binaries with milli-parsec orbital separations are promising nano-Hz GW sources for the rapidly maturing pulsar timing array (PTA) efforts \citep{Liu2023}. These PTA efforts include those by the  International Pulsar Timing Array consortium \cite[IPTA:][]{Falxa2023,Agazie2024} and its constituents, namely the European PTA \cite[EPTA:][]{dcl+2016, epta2023}, the Indian PTA \cite[InPTA:][]{jab+2018,tnr+2022}, the North American Nanohertz Observatory for Gravitational waves \cite[NANOGrav:][]{aaa+2023a,aaa+2023b}, 
the Australia-based Parkes PTA \cite[PPTA:][]{mhb+2013,zrk+2023},  and MeerKAT PTA \citep{msb+2023}. The constituent PTAs of IPTA, namely  NANOGrav, EPTA+InPTA, PPTA and the Chinese PTA in 2023, reported certain compelling evidence for the presence of a stochastic GW background (GWB) in nHz frequencies in their respective data sets \citep{epta+inpta2023a,aaa+2023b,rzs+2023,xcg+2023}.
\par

It is expected that such a GWB could be due to inspiral nHz GWs from an ensemble of SMBH binaries though there are other possible exotic explanations \citep{aaa+2023c, epta+inpta2024}. Additionally,  recent NANOGrav and EPTA+InPTA investigations point towards tentative evidence for GWs from individual SMBH binaries in their latest data sets  \citep{aaa+2023d,epta+inpta2023b}. It is expected that such binaries should allow the IPTA consortium to pursue persistent multi-messenger nano-Hz GW astronomy, especially during the era of 
Square Kilometre Array and Deep Synoptic Array-2000~\citep{SKAPTAa,DSA200,SKAPTAb,Padmanabhan2023}. It may be noted that
SMBH binaries naturally arise from the hierarchical structure formation scenario that involves galaxy mergers \citep{Baugh2006}. Initially, there was a lot of uncertainty about the subsequent evolution of SMBH binaries, especially about their ability to reach GW-driven inspiral phase. However, with later work, both analytic and computer simulations, it has become clear that all binaries formed in this way will merge within a few billion years at most \citep{Begelman1980,Armitage2002,Kulkarni2012,Valtonen2006a,Iwasawa2011,Liao2024}. Unfortunately, electromagnetic observations are only capable of providing potential SMBH binary candidates and 
 unique blazer OJ~287 and PKS 2131-021 are two promising SMBH binary candidates with milli-parsec orbital separations, thanks to decade-long electromagnetic observations \citep{Valtonen2021,Neill2022}
\par 

Thoroughly tested SMBH binary central engine description for OJ~287 is influenced by its century-long light curve, mainly due to the position of this 13-magnitude active galactic nucleus on the ecliptic \citep{Dey2019}. Further, the regular monitoring of OJ~287 in the past decades reveals the presence of quasi-periodic doubly peaked high brightness flares with a deductible period of $\sim 12$ years. The constituent peaks are separated by a few years, and there exists a longer time scale variation in its apparent magnitude with a period of $\sim 60$ years~\citep{Dey2019}. It is possible to explain these observed (and unique) magnitude variations by invoking a central engine description for OJ~287 that involves an SMBH binary\citep{LV96}. This model requires a secondary SMBH that orbits a more massive primary SMBH in a precessing eccentric orbit with a redshifted orbital period of $\sim 12$ years. The inclined secondary BH trajectory ensures that it impacts the accretion disc of the primary twice every orbit leading to the observed Bremsstrahlung impact flares \citep{LV96}. The use of post-Newtonian (PN) accurate orbital trajectory for the secondary SMBH allowed some of us to predict the starting times of the eventually observed impact flares of  2005, 2007, and 2015~\citep{Valtonen2006b,Valtonen2008,Valtonen2016}. 
It may be noted that the PN approximation provides general relativistic corrections to the Newtonian dynamics of an SMBH binary as an expansion in powers of $(v/c)^2\sim GM/c^2r$, where $v$, $M$, $r$ respectively denote the relative velocity, total mass, and the relative separation of the system. For example, the 3PN order provides $(v/c)^6$ order GR corrections to the Newtonian point particle binary dynamics \citep{LB_LR}. Here $c$ is the speed of light. The most up-to-date SMBH binary central engine description for OJ~287 employs PN-accurate orbital dynamics that incorporate 3PN-accurate conservative orbital dynamics associated with non-spinning SMBH binaries, leading order spin-orbit, spin-spin interactions and effects of next-to-next-to-next-to quadrupolar order GW emission~\citep{Dey2018}.
\par 

These considerations allowed us to predict in 2018 that the next impact flare should peak on July 31, 2019, around noon GMT \citep{Dey2018}. The detailed analysis of the multi-epoch Spitzer observations and the predicted similarities between the 2019 and 2007 impact flare lightcurves allowed us to determine that the expected OJ~287's `Eddington flare' arrived only 90 minutes late, within the 4 hours predicted accuracy \citep{Laine2020}. Unfortunately, the retirement of the Spitzer telescope on 30 January 2020 ensured that the predicted  July/August 2022 impact flare was not observable like the Eddington flare \citep{Dey2018}. However, the closely monitored large-amplitude optical intraday flare on 12 November 2021 could be associated with the jet activities from OJ~287's secondary SMBH \citep{V23}. Due to these  multiple successful OJ~287 observational campaigns, launched essentially to verify the predicted occurrences of impact flares,  we have the following measurements for its SMBH binary parameters:
masses of the primary and secondary SMBH are $\sim 18.35 \times 10^9\, M_{\odot} $ and $150 \times 10^6\,M_{\odot}$, respectively while primary SMBH Kerr parameter is $\sim 0.38$ and the orbital eccentricity is $\sim 0.66$. Additionally, both the orbital period (red-shifted) and its time derivative are derived (or estimated) parameters, as emphasized in Table.~2 of \citep{Dey2018}, and their values are $P^{2017}_{\rm obs}=\rm 12.062\pm0.007~yr$ and $\dot{P}_{\rm obs}=-\left(0.00099\pm0.00006\right)$, respectively. Note that the higher-order GW terms decrease the orbital energy loss by $\sim 5$ percent, and therefore it is important that they are included in the orbit solution \citep{Valtonen2018,Dey2018}. Interestingly, the deduced rate of orbital period decay is nine orders of magnitude higher than the observed (measured) rate in PSR 1913+16, as noted in \cite{Dey2018}.
\par

Therefore, it is surprising to see inferences of \cite{Chan2024} where they argued that the SMBH binary orbital dynamics in OJ~287 are affected by dynamical friction induced by certain dark matter (DM) density spikes around its primary SMBH. This conclusion originates from their argument that certain inequality exists between the measured and estimated values of GW energy luminosity, computed using OJ~287's SMBH binary parameters, listed in Table~2 of \cite{Dey2018}. Specifically, \cite{Chan2024} showed that their `estimated' energy loss rate due to GW emission is significantly lower than their `observed' energy loss rate, with a discrepancy of more than $4.3\,\sigma$. This prompted them to employ a dark matter density spike around the primary SMBH, invoking the resulting dynamical friction to provide the extra energy loss rate to account for their inferred discrepancy in OJ~287's observed energy loss rate. These considerations allowed them to estimate spike index  $\gamma_{\it sp}$ whose value turned out to agree with its predicted value based on the adiabatically growing SMBH model~\citep{Gondolo1999,Fields2014}.
\par 

 In what follows,  we take a closer look at the arguments of \cite{Chan2024} and demonstrate that their consistency test strictly works for binary systems like PSR 1913+16 where decades-long accurate timing of $\sim 10,000$ times-of-arrival (ToA) of radio pulses had provided independent measurements for its orbital period and the first-time derivative ($P_b$ \& $\dot P_b$). After that, we provide new estimates for OJ~287's orbital period and its derivative that employ measured values of the independent parameters of SMBH binary central engine description as listed in  Table~2 of \cite{Dey2018}. We also clarify why one should not use the `derived' estimates for OJ~287's orbital period and its derivative as listed in  Table~2 of \cite{Dey2018}. With our new estimates for $P_b$ and  $\dot P_b$, we show that the consistency test of \cite{Chan2024} works for OJ~287 as expected. Interestingly, this test also shows why PN contributions to GW emission are relevant while performing the test. After that, we provide an upper bound for $\gamma_{\it sp}$ by associating the uncertainties in our estimated GW luminosities, influenced by \cite{Horbatsch2012}.


 \section{On the equality between the estimated and measured GW luminosities: The relevance of 
 independently measured parameters }\label{sec2}

We begin by taking a closer look at the consistency test that involves the measured and estimated GW luminosities, as detailed in \cite{Chan2024}. This test, as noted earlier, was pursued to explore if non-GR effects influence the dynamics of OJ~287's SMBH binary.
The test involves computing $\dot{E}_{\rm GW}$ which is the negative of the quadrupolar-order orbital averaged GW luminosity  $<{\cal L}>$~\citep{BS89}, as given by Eq. (2) of \cite{Chan2024}:
\begin{equation}\label{dEdt_GW}  
\dot{E}_{\rm GW}=-\frac{32G^4}{5c^5}\frac{\mu^2M^3}{a^5 (1-e^2)^{7/2}}\left(1+\frac{73}{24}e^2+\frac{37}{96}e^4\right),  
\label{dEdtGW}  
\end{equation}  
where the reduced mass and total mass of the SMBH binary are given by $\mu=m_1 m_2/M$ and $M=m_1+m_2$, respectively, and $m_1$ and $m_2$ stand for the masses of the primary and secondary black holes, respectively. Here $a$ and $e$ denote the semimajor axis and eccentricity parameter, respectively. $G$ denotes the Newtonian gravitational constant and $c$ is the speed of light. \cite{Chan2024} evaluated the above expression by employing the measured parameters of OJ287's SMBH binary model, namely $m_{\rm 1}$, $m_{\rm 2}$, and $e$, as provided in Table 2 of \cite{Dey2018}. However, the value for the orbital semimajor axis $a$ was adopted from\cite{Valtonen1997}, where it was estimated using the redshifted orbital period of $12.07$~yr. 
\par 

In the next step, they obtained the same quantity while using the energy balance argument, which demands that the luminosity of GW emission, namely $<{\cal L}>$, should be balanced by a decrease in the 
Newtonian energy of the compact binary as the rest masses of these compact objects stay constant. The use of Kepler's third law ensures that binary orbital period $P_b$ decreases due to the emission of GWs such that 
\begin{equation}\label{GW_limun}
    \frac{\dot P_b}{P_b}  = +\frac{3}{2\,E\,\mu}\, <{\cal L}>\,,
\end{equation}
where $\dot{P_b}$ stands for the time derivative of $P_b$ and $E$ gives the Newtonian orbital binding energy.
This leads Eq.~(3) of \cite{Chan2024} which reads 
\begin{equation}\label{dEdtorb}
\dot{E}=-\frac{2E\dot{P_b}}{3P_b}\,.
\end{equation}
The consistency test expects numerical estimates for $\dot{E}$ and $\dot{E}_{\rm GW} $ should be identical, provided the parameters that appear on the right-hand side of Eqs.~(\ref{dEdt_GW}) and~(\ref{dEdtorb}) are independently measured. Employing various parameters of OJ~287's SMBH binary description, including redshifted orbital period $P^{2017}_{\rm orb}$ and corresponding decay rate $\dot{P}_{\rm orb}$, listed in Table~2 of \cite{Dey2018} it was argued that there exists a significant $4.3 \sigma$ discrepancy between the numerical values of $\dot{E}$ and $\dot{E}_{\rm GW}$ \citep{Chan2024}. These two estimates turned out to be  $-\left(3.66\pm0.24\right)\times{10}^{41}$~W for $\dot{E}$ and $-\left(2.62\pm0.02\right)\times{10}^{41}$~W for $\dot{E}_{\rm GW}$,
which gave the above discrepancy. This prompted them to explore the possibility of including additional orbital decay mechanisms, such as dynamical friction caused by a dark matter density spike surrounding the primary SMBH, to explain the above-mentioned discrepancy.
\par

A few comments are needed before we delve into the $ {\dot E}_{\rm GW} - {\dot E}$  consistency test of \cite{Chan2024}. First, it is critical to emphasize that the above test works if and only if the parameters used to evaluate Eqs.~(\ref{dEdtGW}) and (\ref{dEdtorb}) are {\it independently measured quantities}. It turns out that this test, in principle, is similar to the way binary pulsars like  PSR B1913+16 are employed to test general relativity (GR) in strong field regimes \citep{Stairs2003}. In the case of binary pulsars, one measures the rate of change of orbital period $\dot P_b$ from the detailed analysis of decades-long ToAs from the constituent pulsar and checks its consistency with GR prediction for $\dot P_b$ that requires independent measurements for the masses of the pulsar and its companion, orbital eccentricity and orbital period. Therefore, performing a similar $ {\dot E}_{\rm GW} - {\dot E}$ consistency test for PSR B1913+16, the most celebrated pulsar binary \citep{TaylorNobel}, would be highly instructive. 
In contrast, a closer inspection of Table 2 of \cite{Dey2018} reveals that it contains OJ~287's measured and estimated parameters categorized as ``independent" and ``derived" quantities, respectively. Unfortunately,  \cite{Chan2024} did not differentiate between these two types of parameters while employing them to perform the above consistency test for the SMBH binary in OJ~287. An additional point is that \cite{Dey2018} did not provide any direct measurement of $a$, OJ~287's semi-major axis. Therefore, \cite{Chan2024} relied on Kepler's third law, i.e., $P_b \propto a^{3/2} $ and incorporated derived values of $a$ from \cite{Valtonen1997}. In our opinion, it is not necessary to invoke semi-major axis $a$, especially when we have estimates for a gauge invariant quantity like $P_b$~\citep{DS88}. This is another reason to take a closer look at their consistency test for OJ~287.
\par 

To clarify that $ {\dot E}_{\rm GW} - {\dot E}$ consistency test works {\it if and only if} the employed parameters are independently measured, we consider the observations of the first relativistic binary pulsar PSR~B1913+16 \citep{Hulse1975}. By analyzing 9257 precise TOAs collected over 35 years using the Arecibo Observatory, \cite{Weisberg2016} showed that the observed orbital period decay for PSR B1913+16 closely matched with the associated GR-prediction, with a discrepancy of less than $\sim 1\sigma$. To ensure that we employ only the {\it measured parameters} that arise from the many decades-long precise timing observations of PSR~B1913+16, it is imperative to use the following  quadrupolar order expression for $ \dot{E}_{\rm GW}$, extracted from Equation~(4.20) in \cite{BS89}:
\begin{equation} \label{psr_Edot_GW}
      \dot{E}_{\rm GW} = -\frac{32}{5} \frac{\eta^2\left( G\,M\, n/c^3\right)^{\frac{10}{3}}}{\left(1-e^2\right)^{\frac{7}{2}}} \left(1+\frac{73}{24}e^2+\frac{37}{96}e^4\right),
\end{equation}
where $\eta = m_1 m_2/\left(m_1+m_2\right)^2$ is the symmetric mass ratio, and $n = 2\,\pi/P_b$ is the gauge invariant mean motion in $s^{-1}$. We have used $T_\odot  = G\,M_\odot/c^3 = 4.925,490,947 \mu s$ while evaluating $ (G\,M\,n/c^3)$ where $M_\odot$ stands for the solar mass. Additionally, we have scaled the above expression by $c^5/G$ to deal with only a dimensionless expression, and this also takes care of the fact that $G$ is a poorly measured quantity \citep{Xue2020}. We now express the time derivative of the orbital (binding) energy, given by Eq.~(\ref{dEdtorb}), in terms of the orbital period $P_b$ and its time derivative $\dot{P}_b$, and it reads 
\begin{equation} \label{psr_Edot_orb}
    \dot{E} = \frac{\eta\left( G\,M\, n/c^3\right)^{\frac{5}{3}} \dot{P_b} }{6\pi}.
\end{equation}  
This expression is also scaled by $c^5/G$ so that we deal only with dimensionless expressions. 
\par 

We now have all the inputs to pursue the $ {\dot E}_{\rm GW} - {\dot E}$ consistency test with the help of independently measured parameters of an inspiraling compact binary.
For pursuing the consistency test, we evaluate Equation~(\ref{psr_Edot_GW}) using the measured values of $P_b=0.322997448918(3)$~day, $e=0.6171340(4)$, $m_1=1.438 \pm 0.001~\rm M_\odot$, and $m_2=1.390 \pm 0.001~\rm M_\odot$, available in Table.~2 of \cite{Weisberg2016}. This results in the dimensionless  $ {\dot E}_{\rm GW} = -\left(2.14009\pm0.00357\right)\times {10}^{-28}$  estimate for PSR~B1913+16. We would like to emphasize that the above-measured values for the binary pulsar's orbital elements and parameters arise from the precise modeling of various relativistic effects that influence the ToAs of PSR~B1913+16's radio pulses \citep{Weisberg2016}. These delays include the Roemer delay, caused by the motion of the pulsar in its orbit; the Shapiro delay, resulting from the curvature of spacetime near the companion star; and the Einstein delay, which accounts for gravitational time dilation and the pulsar's varying orbital velocity while incorporating post-Keplerian effects like the advance of periastron \citep{Weisberg2016,Stairs2003}.
\par

We now calculate the dimensionless value of $ {\dot E} $ using Equation~(\ref{psr_Edot_orb}). Employing $ \dot P_b = -\left(2.398\pm0.004\right)\times{10}^{-12}$ for PSR~B1913+16, as obtained through Equation (15) of \cite{Weisberg2016}, and the resulting expression evaluates to $-\left(2.13643\pm0.00398\right)\times{10}^{-28}$. A close inspection of these two estimates reveals that the $ {\dot E}_{\rm GW} - {\dot E}$ consistency test works for PSR~B1913+16. It is important to note that the parameter ${\dot P_b}$ enters as an independent parameter in the Kepler Equation, expressed in terms of proper time coordinates, in the heavily used Damour-Deruelle timing model \citep{DD86,Stairs2003} and it reads 
\begin{equation}\label{eq6}  
    u - e \sin u =  2\,\pi \biggl \{ 
    \frac{(t - t_0)}{P_b}  - \frac{\dot P_b}{2} \biggl (\frac{(t - t_0)}{P_b} \biggr )^2 
\biggr \}\,,
\end{equation}  
where $u$ is the eccentric anomaly, and $t_0$ is the reference time for periastron passage. The use of the Damour-Deruelle timing formula ensures that the long-term timing of relativistic binary pulsars regularly leads to independently measured values for  $P_b$, $\dot{P}_b, e$, along with the masses $m_1$ and $m_2$~\citep{Stairs2003,DT92}. Influenced by \cite{Weisberg2016}, we take the ratio of $ {\dot E}_{\rm GW}$ and $ {\dot E}$ for the binary pulsar and we get its value to be $0.9983 \pm 0.0025$ which quantifies the consistency test of \cite{Chan2024} for PSR B1913+16.
\par

A few comments are in order. In the case of binary pulsars, it is customary to obtain estimates for the quadrupolar order $\dot P_b$ expression, given by Eq.~(22) in \citep{Weisberg2016} and compare it with the measured intrinsic $\dot P_b$ values, given by Eq.~(15) in \citep{Weisberg2016}. For PSR B1913+16, the ratio between $ {\dot P_b}^{\rm intr}$ and $ {\dot P_b}^{\rm GR}$ turned out to be $0.9983 \pm 0.0016$ as evident from Eq.~(23) in \citep{Weisberg2016}. This result implies that the binary pulsar is losing energy to GWs within $\sim 1\sigma$ of the rate predicted by GR. This is essentially the reason why the long-term timing of PSR B1913+16 provided the first indirect evidence for the existence of GWs \citep{TaylorNobel}. In contrast, the $ {\dot E}_{\rm GW} - {\dot E}$ consistency test, as demonstrated by \cite{Chan2024}, is rather convoluted from the PSR B1913+16 perspective. This is because it requires us to employ $P_b, \dot{P}_b$, $m_1, m_2$ and $e$ values while evaluating both the expressions and thereby mixing measured quantities that arise from the conservative and dissipative aspects of  PN-accurate orbital dynamics of the compact binary present in PSR B1913+16 \cite{DT92}.
\par

It should be obvious from these discussions that independent measurements of orbital elements and parameters are critical for performing the $ {\dot E}_{\rm GW} - {\dot E}$ consistency test. In contrast to relativistic binary pulsars, both $P_b$  and $\dot P_b$ are not observationally measurable quantities in our SMBH binary description for OJ~287 \citep{Dey2018}. This is mainly because we only have observational data for two epochs per orbit that are associated with secondary SMBH crossings for determining various astrophysical and relativistic aspects of OJ~287's SMBH binary, listed as `independent' parameters in Table 2 of~\cite{Dey2018}. Further, a pair of these epochs are not separated by exactly one orbital period due to substantial relativistic precession. These considerations prompted some of us to provide rough estimates for both the orbital period and its time derivative using the complete orbit solution/trajectory, and they were displayed as `derived' estimates in Table 2 of \cite{Dey2018}. These rough estimates were displayed essentially to show the highly relativistic nature of OJ~287's SMBH binary compared to relativistic binary pulsars. In Table 2 of \cite{Dey2018}, we extracted the orbital period value from the difference in two subsequent epochs of  secondary SMBH apastron passages, namely between 2013 and 2025 apocenter times.  The epochs of apastron passages  rather than pericenter passages were chosen due to the comparatively lower speed of the secondary at these epochs which allowed dense sampling of the trajectory numerically. After that, we estimated $\dot P_b$ using the differences in the $P_b$ values between 1901 and 2021 and dividing them by the elapsed time of 120 years. We would like to emphasize that these derived quantities played no role 
in the prediction of 2019 impact flare though their accuracies are influenced by the orbital trajectory constructed with the help of Table.~2  of \cite{Dey2018}. 
\par

It is indeed possible to obtain accurate estimates for $P_b$, and its time-derivative using the `independent' parameters in Table 2 of \cite{Dey2018} with the help of the following steps. First, we compute $P_b$ estimate in the rest frame of the SMBH binary by employing $\Delta \Phi = 2\pi k$ relation, where $\Delta \Phi$ represents the precession rate of the major axis per period, and $k$ denotes the fractional rate of advance of the periastron, and we use the 3PN-accurate expression for $k$ available in \cite{Gopu2006} (see also Equation A1 in~\cite{Dey2018}). Using measured parameters from Table 2 in \cite{Dey2018}, we calculate the non-redshifted orbital period as $P_b = 9.231\pm0.034$~yr. Including the redshift effect yields orbital period as $P_{\rm obs} = 12.056\pm0.044$~yr, which is not very different from the listed value of  $P^{2017}_{\rm orb} = 12.062\pm0.007$~yr in \cite{Dey2018}. For obtaining an accurate estimate for $\dot P_b$, we invoke Equation (2.8) from \cite{BS89} and the relevant measured parameters in Table 2 of \cite{Dey2018}, and it leads to $\dot{P}_b = -(0.000515 \pm 0.000006)$. We are now in a position to perform the $ {\dot E}_{\rm GW} - {\dot E}$ consistency test. Substituting the above-derived values into Equations (\ref{psr_Edot_GW}) and (\ref{psr_Edot_orb}), we obtain $\dot{E}_{\rm GW} = -(6.782\pm0.111)\times 10^{-12}$ and $\dot{E} = -(6.782\pm0.520)\times 10^{-12}$ in dimensionless units. It should be obvious that there is no statistically significant difference between ${\dot E}_{\rm GW} $ and $ {\dot E}$ as their central values are essentially identical. These results confirm that the ${\dot E}_{\rm GW} - {\dot E}$ consistency test holds well for OJ~287, provided we employ only the measured (`independent') parameters associated with its SMBH binary central engine description. The use of general relativistic inputs and observationally measured parameters of OJ~287, based on its SMBH binary central engine description, in the above consistency test show that the orbital dynamics of SMBH binary in OJ~287 accurately follow general relativity. We want to note in passing that it is customary to use the dominant PN contribution to $k$ for constraining the total mass of binary pulsars while the quadrupolar order $\dot P_b$ expression is employed to estimate the expected rate of orbital decay in newly discovered binary pulsars \citep{Barr24}.
\par

To validate our above-mentioned quadrupolar order approach, we conducted a 1PN-accurate ${\dot E}_{\rm GW} - {\dot E}$ consistency test using expressions from \cite{BS89}, yielding $\dot{P}_b = -(0.000684 \pm 0.000009)$. This analysis revealed a good agreement between 1PN-accurate ${\dot E}_{\rm GW}$ and ${\dot E}$ estimates, with a discrepancy of only $\sim 0.41~\sigma$. Specifically, the 1PN-accurate forms of Eqs.~(\ref{psr_Edot_GW}) and (\ref{psr_Edot_orb}) give $\dot{E}_{\rm GW} = -\left(9.268 \pm 0.161\right)\times{10}^{-12}$ and $\dot{E} = -\left(9.355 \pm 0.136\right)\times{10}^{-12}$, respectively. We pursued the above computations as the orbital dynamics of OJ~287 employ PN-accurate equations of motion that incorporated GW emission effects beyond the quadrupolar order~\citep{Dey2018}. It is worthwhile to note that 
 an additional subtlety was not addressed in \citep{Chan2024}. It is desirable to employ the non-redshifted orbital period in theoretical calculations for $\dot{E}$ and $\dot{E}_{\rm GW}$
 as these quantities are usually provided in the rest frame of a compact binary system. However, \cite{Chan2024} used the redshifted $P^{2017}_{\rm orb}$ and the corresponding $\dot{P}_{\rm orb}$, as listed in Table 2 of \cite{Dey2018}, in their the ${\dot E}_{\rm GW} - {\dot E}$ consistency test and their reported $4.3\sigma$ discrepancy. Interestingly, our quadrupolar and 1PN-accurate consistency tests required us to use the non-redshifted orbital period. These considerations suggest that the reported evidence for a dark matter density spike around the primary SMBH in OJ~287 and the derived spike index $\gamma_{\rm sp}$ require further scrutiny and revision. In what follows, we provide certain upper bounds for $ \gamma_{\rm sp} $, derived under the revised framework.

\section{Constraining $\gamma_{\rm sp}$ Using Timing Uncertainties in OJ~287's Impact Flare Observations }

Our approach to constrain $\gamma_{sp}$ is influenced by an effort by \cite{Horbatsch2012} who constrained the instantaneous time-variation of `sufficiently light'  scalar fields using inputs from OJ~287's  SMBH binary description. This effort relies on an earlier result that showed that a BH should acquire `scalar hair', provided the underlying scalar field is slowly time-dependent far from the BH \citep{TJ99}. Invoking the above result, it was argued that an orbiting pair of BHs can radiate dipole radiation, provided the two BHs have different masses \citep{Horbatsch2012}.  Additionally, \cite{Horbatsch2012} computed an appropriate formula for the emitted power from the time-varying dipole moment, induced by the time-dependent scalar hair that resides on the constituent BHs in an inspiraling BH binary. It turns out that the ratio of dipolar to quadrupolar GW luminosities essentially depends on the frequency $\mu'$, which characterizes the variation of the scalar field far from the black hole binary and it requires good estimates for various orbital elements and parameters like $m_1,m_2, e$ and $P_b$\citep{Horbatsch2012}. This prompted \cite{Horbatsch2012} to equate the above ratio, computed for OJ~287's SMBH binary description as given in
\cite{Valtonen2008}, to $< 0.06$.
This fraction, as expected, essentially quantifies the uncertainty in predicting the arrival time of the observed 2007 impact flare while using 2.5PN-accurate orbital dynamics of non-spinning SMBH binaries~\citep{Valtonen2008}. The resulting upper bound on the time-variation of scalar fields turned out to be  $ < (16 \, \text{days})^{-1}$. In what follows, we adopt a similar approach to constrain $\gamma_{sp}$.
\par

 Recall that \cite{Chan2024} introduced an additional mechanism involving energy dissipation through dynamical friction caused by a dark matter density spike surrounding the primary SMBH in OJ~287. This was, as noted earlier, influenced by their contestable inference that OJ~287's SMBH binary central engine description is inconsistent with their $ {\dot E}_{\rm GW} - {\dot E}$ consistency test. To model the dark matter density near the primary SMBH, \cite{Chan2024} employed the following spike profile:
\begin{equation}\label{DM_den}
\rho_{\rm DM}=\left\{
\begin{array}{ll}
0 & {\rm for }\,\,\, r\le 2R_s, \\
\rho_{\rm sp} \left(1-\frac{2R_s}{r} \right)^3 \left(\frac{r}{r_{\rm sp}} \right)^{-\gamma_{\rm sp}} & {\rm for }\,\,\, 2R_s <r \le r_{\rm sp}, \\
\frac{\rho_sr_s}{r} & {\rm for}\,\,\, r_{\rm sp}<r \ll r_s, \\
\end{array}
\right.
\end{equation}
where $r_s$ and $\rho_S$ represent the scale radius and scale density, respectively, while the term $R_s=2Gm_1/c^2$ stands for the Schwarzschild radius while  $r_{sp}$ is the radius of spike region and $\gamma_{\rm sp}$  is the spike index characterizing the DM density gradient within the spike region. The resulting orbital averaged  energy loss rate  due to dynamical friction induced by the DM density spike surrounding the central SMBH reads~\citep{Yue2019}
\begin{eqnarray}~\label{Edot_DF}
\dot{E}_{\rm DF}&=&-2G^{\frac32}\mu^2\rho_{\rm sp}r_{\rm sp}^{\gamma_{\rm sp}}(1-e^2)^{\frac32}\ln\Lambda  \int_0^{2\pi} \frac{[1+e\cos(1-\alpha)\phi]^{\gamma_{\rm sp}-2}[p-2R_s(1+e\cos(1-\alpha)\phi)]^3}
{p^{\gamma_{\rm sp}+\frac52}m_1^{\frac{1}{2}}[1+2e\cos(1-\alpha)\phi+e^2]^{\frac12}} d \phi,
\end{eqnarray}
where $\ln \Lambda \approx \ln \sqrt{m_1/m_2}$ is the Coulomb logarithm, $p$ is the semi-latus rectum, $\alpha$ denotes the precession phase angle, and $\phi$ is the orbital phase. If the dark matter density is extremely high near the primary SMBH, dynamical friction could substantially alter the inspiral dynamics of OJ~287's SMBH binary. Therefore, \cite{Chan2024} argued that the above contribution should be added to $ \dot{E}_{\rm GW}$ while performing the above consistency test for OJ~287. In other words, 
\cite{Chan2024} argued that it was 
the neglect of $ \dot{E}_{\rm DF}$  contribution that caused the failure of  $ {\dot E}_{\rm GW} - {\dot E}$ consistency test for OJ~287 while using the measured and estimated parameters listed in Table.~2 of \cite{Dey2018}. This prompted \cite{Chan2024}  to equate $ {\dot E}_{\rm GW} + \dot{E}_{\rm DF}$, given by Eqs.~(\ref{dEdtGW}) and (\ref{Edot_DF}) to $ \dot{E} $ expression, given by Eq~\ref{dEdtorb},  while using various parameters listed in Table.~2 of \cite{Dey2018} 
and value of $a$ as obtained by \cite{Valtonen1997}.
This resulted in Figure~1 of \cite{Chan2024}, and it should be obvious that they were able to estimate accurately a value for the dark matter density spike $\gamma_{\rm sp} = 2.351^{+0.032}_{-0.045}$. It turned out that this value aligns closely with the canonical model prediction of $\gamma_{\rm sp} = 2.333$ \citep{Gondolo1999,Lacroix2018}.
In our opinion, this estimate requires further scrutiny due to our way of 
showing the $ {\dot E}_{\rm GW} - {\dot E}$ consistency for OJ~287 as 
discussed in Section~\ref{sec2}.

\par


\begin{figure}[!htpb]
\centering
\includegraphics[width=0.50\textwidth]{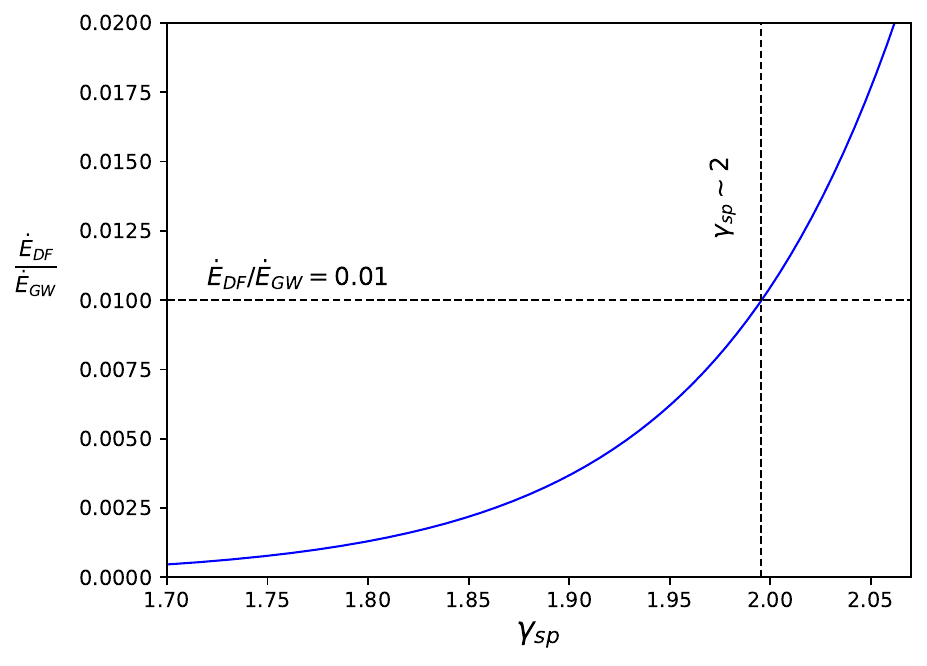}
\includegraphics[width=0.48\textwidth]{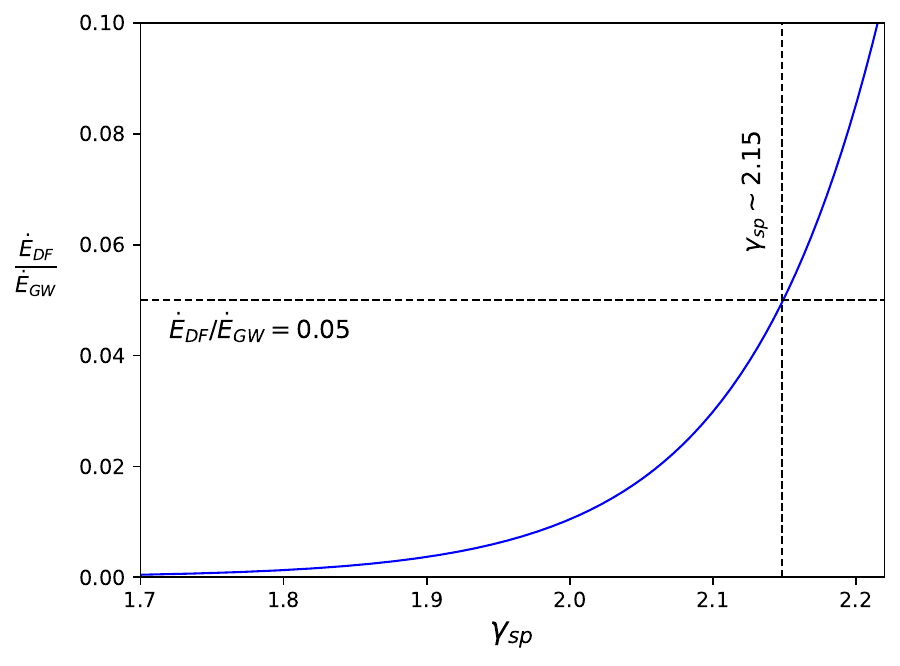}
\caption{
Plots for the variation of the ratio $\dot{E}_{\rm DF} / \dot{E}_{\rm GW}$ as a function of $\gamma_{\rm sp}$ while using SMBH binary description of OJ~287 as given in Table 2 of \cite{Dey2018}. The uncertainty levels associated with the observed 2015 and 2019 impact flares in our model lead to upper bounds marked by the dashed lines, influenced by the approach of~\cite{Horbatsch2012}. Both these upper bounds are inconsistent with the estimated value of $\gamma_{\rm sp}$ by \cite{Chan2024}.
} \label{E_ratio_fig}
\end{figure}


Our approach to constrain $\gamma_{\rm sp}$, as noted earlier, is influenced by \cite{Horbatsch2012} and therefore, it involves computing first the ratio $\dot{E}_{\rm DF} / \dot{E}_{\rm GW}$, given by Equations~(\ref{psr_Edot_GW}) and (\ref{Edot_DF}). Thereafter, we equate the ratio to the uncertainties in our ability to predict the observed 2015 and 2019 impact flares from OJ~287. The 2019 outburst, originating near the disk's pericenter, turned out to be fairly identical to the 2007 outburst in our SMBH binary central engine description for OJ~287 \citep{Dey2018}. The eventual observation of the 2019 outburst with a light curve similar to the one associated with the 2007 impact flare allowed us to provide consistency in our observations with its GR prediction to within $1\%$\citep{Laine2020}.
In contrast, the 2015 flare, originating closer to the apocenter with timing consistency within $5\%$ of GR predictions~\citep{Valtonen2016}. Therefore, we let the above ratio be 1\%  and 5\%  to obtain upper bounds on  $\gamma_{\rm sp}$. We obtain upper bound of $\gamma_{\rm sp}$ as $\sim 2$ and $2.15$ for uncertainties of $1\%$ and $5\%$, respectively, by invoking the ratio $\dot{E}_{\rm DF} / \dot{E}_{\rm GW}$. In other words, we argue that uncertainties in predicting the observed impact flares may be associated with OJ~287's orbital damping induced by dynamical friction associated with a dark matter density spike around its primary SMBH. Additionally, we used the non-redshifted $P_b$ instead of the semimajor axis to compute $\dot{E}_{\rm DF}$, as given by Eq.~(\ref{Edot_DF}). The results are presented in Figure~\ref{E_ratio_fig}. It should be obvious that it is not possible for us to estimate an accurate value for $\gamma_{\rm sp}$ but provide only interesting upper bounds. This is not surprising as we do not follow the arguments of \cite{Chan2024} that the $\dot{E}_{\rm GW}$ value needs to be supplemented by the $\dot{E}_{\rm DF}$ contribution to account for their inferred value for  $ \dot E$. Moreover, since the $ {\dot E}_{\rm GW} - {\dot E}$ consistency test performs well for SMBH binary systems like OJ~287, it supports the presence of `sufficiently light' scalar fields around OJ~287, as argued by \cite{Horbatsch2012}. This method provides a plausible approach to constrain the upper bound of $\gamma_{\rm sp}$ using the ratio $\dot{E}_{\rm DF} / \dot{E}_{\rm GW}$ while accounting for the uncertainties in predicting the observed impact flares of OJ~287.


\section{Conclusions and Discussions}
We provide a way to pursue the $ {\dot E}_{\rm GW} - {\dot E}$ consistency test for OJ~287, proposed by \cite{Chan2024}, while employing independently measured parameters from Table~2 of \cite{Dey2018} for describing OJ~287's SMBH binary central engine. This prompted us to probe the reported evidence for a dark matter density spike around the primary SMBH in OJ 287 and their way of estimating the spike index $\gamma_{\rm sp}$. After that, we provided upper bounds for $\gamma_{\rm sp} $ by employing the uncertainties in the observed 2015 and 2019 impact flares to their GR-based predictions, influenced by \cite{Horbatsch2012}. The resulting upper bounds $\gamma_{\rm sp}$ turned out to be $\sim 2$ and $2.15$ when we let the ratio $\dot{E}_{\rm DF} / \dot{E}_{\rm GW}$ to be $0.01$ and $ 0.05$, respectively.
\par

The above upper bounds may have implications for a recent interesting effort that placed an upper bound on the DM spike mass surrounding the primary SMBH of OJ~287~\citep{Alachkar2023}. This detailed analysis puts the above mass to be less than $3\%$ of the mass of the primary SMBH, assuming $\gamma_{\rm sp} \sim 2.333$~\citep{Lacroix2018}. Additionally, it would be worthwhile to explore how these upper bounds on $\gamma_{\rm sp}$ influence the constraints on the dark matter spike mass around the primary SMBH in OJ287, as discussed in \cite{Alachkar2023}, potentially advancing our understanding of DM profiles in such systems.

\begin{acknowledgments}
We thank Gonzalo Alonso Alvarez for helpful discussions and suggestions.
DD acknowledges the Department of Atomic Energy, Government of India's support through `Apex Project-Advance Research and Education in Mathematical Sciences' at The Institute of Mathematical Sciences. AG acknowledges the support of the Department of Atomic Energy, Government of India, under project identification $\mathrm{\# RTI\ 4002}$. We sincerely thank the anonymous referee for their insightful comments and valuable suggestions.
\end{acknowledgments}

\end{document}